\documentclass[aps,a4paper,10pt,twocolumn,nofootinbib]{revtex4} 
\usepackage{datetime}
\usepackage{amsmath}
\usepackage{amsfonts}
\usepackage{amsfonts}
\usepackage{eufrak}
\usepackage{mathrsfs}
\usepackage[mathscr]{euscript}
\usepackage[dvips]{graphicx}
\usepackage{fancyhdr}
\usepackage{colordvi}
\usepackage{hyperref} 

\def\overstrike#1#2{{\setbox0\hbox{$#2$}\hbox to \wd0{\hss
    $#1$\hss}\kern-\wd0\box0}}

\newdateformat{yymmdddate}{\THEYEAR/\twodigit{\THEMONTH}/\twodigit{\THEDAY}}

\begin{document}
\title{Drift, diffusion, and third order derivatives in Fokker-Planck equations: one specific case}
\author{Paul Kinsler}
\email{Dr.Paul.Kinsler@physics.org}
\affiliation{
  Blackett Laboratory, Imperial College,
  Prince Consort Road,
  London SW7 2AZ, 
  United Kingdom.
}

\lhead{\includegraphics[height=5mm,angle=0]{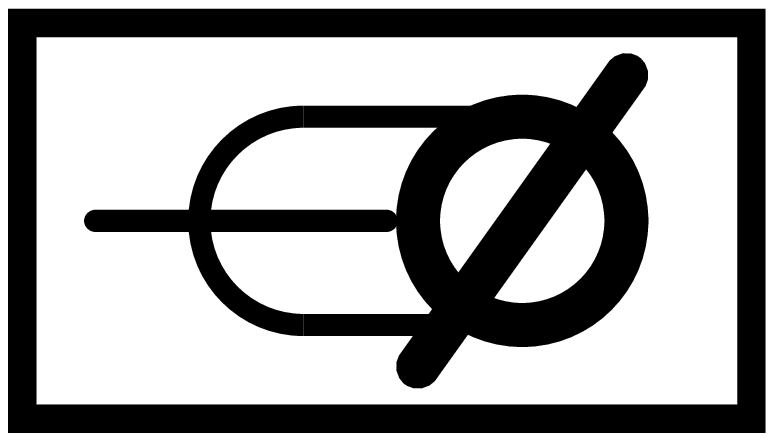}~~FPE3RD}
\chead{Third derivatives and FPE's}
\rhead{
\href{mailto:Dr.Paul.Kinsler@physics.org}{Dr.Paul.Kinsler@physics.org}\\
\href{http://www.kinsler.org/physics/}{http://www.kinsler.org/physics/}
}

\begin{abstract}

I present a case where there is an exact re-interpretation 
 for the third order derivative term in a Fokker-Planck equation, 
 purely in terms of ordinary drift and diffusion.

\end{abstract}

\date{\today}
\maketitle
\thispagestyle{fancy}

%

%
\section{Introduction}\label{S-intro}

There are many situations in optics where we would like to 
 treat system dynamics using those methods developed for
 Fokker-Planck equations, 
 or their stochastic analogues.
Some notable cases are for the 
 quantum-optical parametric oscillator \cite{Kinsler-D-1991pra}
 and the cold-atom Gross-Pitaevski equation 
 for Bose-Einstein (and other) 
 condensates \cite{Sinatra-LC-2002,Gardiner-AF-2002jpb,Cherroret-W-2011pre}
However, 
 some (many) systems in fact give rise to partial differential equations
 containing extra terms, 
 notably derivative terms of higher order than second.
Fortunately,
 there are often grounds for considering 
 such terms to negligible, 
 so they are neglected (``truncated'') -- 
 thus giving rise to the so-called ``truncated Wigner'' phase space
 \cite{Kinsler-D-1991pra,Sinatra-LC-2002}
 descriptions of quantum optical dynamics.
Nevertheless, 
 just because a term is small, 
 that does not preclude it gradually accumulating 
 and so providing a significant distortion.
An ideal test bed for such comparisons
 is the second-order nonlinear interection
 present in the parametric ocsillator ..... 
 to the dynamics\cite{Drummond-K-1989pra,Kinsler-D-1995,Kinsler-1996}, 
 so attempts have been made to estimate these effects.
In this note I show that an exact comparison can be made
 between the distribution function of a system 
 follwing a purely third order differential equation
 and one following a purely second order (Fokker-Plank) form
 with modified drift and diffusion terms.

%
\section{Fokker-Planck equations}\label{S-FPE}

Fokker-Planck equations (FPE's) are usually of the form 
 \cite{RiskenFPE,GardinerHSM,ArnoldSDETA}
~ 
\begin{align}
  \partial_t P(x;t)
&=
  \left[
    \partial_x A(x)
   +
    \partial_x^2 B(x)
  \right]
  P(x;t)
\\
&=
  \partial_x
  \left[
    A(x)
   +
    \partial_x B(x)
  \right]
  P(x;t)
.
\label{eqn-intro-FPE2}
\end{align}
This partial differential equation describes how the 
 probability distribution function $P$ evolves in space ($x$)
 as a function of time ($t$).
The form gven here is one dimensional, 
 but of course multi-dimensional generalizations also exist.
Here $A(x)$ is a \emph{drift} term
 causing probability to flow deterministically, 
 and $B(x)$ is a \emph{diffusion} term
 causing probability to spread away from some give point.

However, 
 in some contexts we can generate FPE-like equations that have additional 
 third order derivative terms.
An example is when using the Wigner representation 
 to derive a FPE for the optical parametic oscillator\cite{Kinsler-D-1991pra}.
Such a FPE could be of the simple form
~ 
\begin{align}
  \partial_t P(x;t)
&=
  \partial_x
  \left[
    A(x)
   +
    \partial_x B(x)
   +
    \partial_x^2 C(x)
  \right]
  P(x;t)
.
\label{eqn-intro-FPE3}
\end{align}
It is not uncommon in simple quantum optics treatments to 
 simply truncate (i.e. neglect) these third order terms.
Even if we chose to take this step, 
 we would often like to know what 
 perturbations these terms would have made to our 
 approximate solution.
So, 
 what \emph{interpretation} should (or can) 
 we place on the term $C(x)$?
This is a hard problem in general, 
 but I show here that for a very simple and specific case
 based on Hermite polynomials, 
 an exact identification can be made.
This then can motivate physical intuition 
 as to the likely effects of such third-order terms 
 in more general cases.

%
\section{Hermite polynomials}\label{S-hermite}

Hermite polynomials are a set of orthogonal polynomials, 
 and, 
 as such, 
 can be used to describe any arbitrary function 
 using the form
~ 
\begin{align}
  P(x;t)
&=
  \sum_n
    a_n(t) H_n(x)
.
\label{eqn-hermite-P}
\end{align}

For the purpose of this short note, 
 they have some useful recurrence properties, 
~ 
\begin{align}
  H_{n+1}(x)
&=
  2 x H_{n}(x)
 -
  2 n H_{n-1}(x)
\label{eqn-hermite-recurr1}
,
\\
  \partial_x
  H_{n}(x)
&=
  2 n H_{n-1}(x)
\label{eqn-hermite-recurr2}
, 
\\
\textrm{hence}
~~~~
~~~~
  H_{n+1}(x)
&=
  2 x H_{n}(x)
 -
  \partial_x
  H_{n}(x)
.
\end{align}
When used appropriately, 
 these recurrence properties enable us to convert 
 a FPE containing third-order derivative terms 
 into a form containing only second order ones \cite{WfmMathWorld-HermitePolynomial}.

%
\section{Transforming the third derivatives}\label{S-thirdderivs}

Using eqn. (\ref{eqn-hermite-P}) we can decompose the full FPE
as follows 
~ 
\begin{align}
  \sum_n
    a_n(t) H_n(x)
&= 
  \partial_x
  \left[
    A(x)
   +
    \partial_x B(x)
   +
    \partial_x^2 C(x)
  \right]
\nonumber
\\
&
~~~~
~~~~
  \times
  \sum_n
    a_n(t) H_n(x)
\label{eqn-thirdderivs-hermiteFPE}
\end{align}

Note that since there is no cross-coupling between the $H_n(x)$ 
 components, 
 and because the $H_n(x)$ are orthogonal, 
 we can write 
~
\begin{align}
  \partial_t a_n(t) H_{n}(x)
&=
  \partial_x
  \left[
    A(x)
   +
    \partial_x B(x)
   +
    \partial_x^2 C(x)
  \right]
  a_n(t)
  H_{n}(x)
\label{eqn-thirdderivs-hermiteFPE-n}
\end{align}

%
\subsection{A single $H_{n}$ component}\label{S-Hn}

I now 
 assume $C(x)=C$ is a constant\footnote{??as 
  it is for the truncated-Wigner dpo case?},
 and consider the RH derivative part
 of eqn. (\ref{eqn-thirdderivs-hermiteFPE-n}).
Considering just one component $H_n$ of the Hermite decomposition
 of $P$, 
 we have a RHS from \eqref{eqn-thirdderivs-hermiteFPE-n} of
~ 
\begin{align}
  \partial_t a_n(t) H_{n}(x)
&=
  \partial_x
  \left[
    A(x)
   +
    \partial_x B(x)
   +
    \partial_x^2 C
  \right]
  H_{n}(x)
\\
&=
    \partial_x
    A(x) H_{n}(x)
   +
    \partial_x^2 B(x) H_{n}(x)
\nonumber
\\
& \qquad\quad
   +
  \partial_x^2
      C
      \left(
        2 x H_{n} - H_{n+1}
      \right)
\\
&=
    \partial_x
    A(x) H_{n}(x)
   +
    \partial_x^2 B(x) H_{n}(x)
\nonumber
\\
& \qquad\quad
   +
  \partial_x^2
      C
        2 x H_{n}
   +
  \partial_x^2
      C
      H_{n+1}
\\
&=
    \partial_x
    A(x) H_{n}(x)
   +
    \partial_x^2 B(x) H_{n}(x)
\nonumber
\\
& \qquad\quad
   +
  \partial_x^2
      C
        2 x H_{n}
   +
  \partial_x
      C
      2 (n+1)
      H_{n}
\\
&=
    \partial_x
  \left[
    A(x) 
   +
      2 (n+1)
      C
  \right]
    H_{n}(x)
\nonumber
\\
& \qquad\quad
   +
    \partial_x^2
  \left[
    B(x) 
   +
    2x
  \right]
    H_{n}(x)
\\
&=
    \partial_x
  \left[
    A'(x) 
   +
    \partial_x
    B'(x) 
  \right]
    H_{n}(x)
,
\end{align}
where
~ 
\begin{align}
  A'_n(x)
&=
      A(x) 
   +
      2 (n+1)
      C
\\
  B'(x)
&=
    B(x) 
   +
    2 x C
.
\end{align}

So for a distribution function $P$ decomposed
 into Hermite polynomials $H_n$, 
 a third-order derivative term 
 with a constant prefactor $C$
 has two effects:

\begin{enumerate}

\item
The first, 
 and presumably most important feature
 is that the drift term $A$ gains
 a constant positive contribution.
We would typically expect
 such a directional property, 
 since $\partial_x^3$ is clearly antisymmetric in nature.
The strength of this effective drift is dependent on $n$, 
 the polynomial order, 
 so that fine $x$ structure in $P$ 
 (requiring higher order contributions)
 drifts faster than coarse features;

\item
The second, 
 more minor effect is to add an antisymmetric 
 adjustment to the diffusion.

\end{enumerate}

The transformed counterpart to eqn. (\ref{eqn-thirdderivs-hermiteFPE-n}) is
~
\begin{align}
  \partial_t a_n(t) H_{n}(x)
&=
  \partial_x
  \left[
    A'_n(x)
   +
    \partial_x B'(x)
  \right]
  a_n(t)
  H_{n}(x)
.
\label{eqn-thirdderivs-hermiteFPE-n-trans}
\end{align}

This therefore is a nice specific example
 where we can re-cast a third order derivatve term 
 into the readily understood first order (drift)
 and second order (diffusion) terms.
If implemented in
 some numerical scheme, 
 it would require repetition of the following steps:

\begin{description}

\item[(a)] 
 the distribution $P(t)$ to be decomposed 
 into $H_n$ with weights $a_n(t)$,

\item[(b)] 
 each $H_n$ to evolve away from an exact Hermite polynomial
 under the influence of the drift and diffusion
 for some suitably small time interval $\delta t$, 

\item[(c)] 
 an evolved distribution $P(t+\delta t)$ to be calculated.

\end{description}

Whether or not this is useful in practise is left as an exercise
 for those dealing with such situations.
However, 
 even if such an implementation is not done, 
 and the third-order derivatives are simple truncated as usual, 
 the behaviour of $A'_n$ can be used to put constraints
 on the size of spatial features on
 distributions $P$ evolved using a truncated FPE.
Thus it might at least be put to use as an
 intermittently applied test on the validity
 of a simulation of a truncated FPE model.

%
\subsection{Rearranged $H_{n}$ component}\label{S-Hn-B}

We might attempt an alternative strategy
 that tries to avoid the requirement 
 of decomposing the distribution function $P$.
However, 
 the one given below will not work, 
 but for the sake of completeness I give it here.

Firstly, 
 assume $C(x)=C$ is a constant
 and consider the RH derivative part
 of eqn. (\ref{eqn-thirdderivs-hermiteFPE-n}), 
~ 
\begin{align}
  \partial_t a_n(t) H_{n}(x)
&=
    \partial_x
    A(x) H_{n}(x) a_n(t)
\nonumber
\\
& \qquad\quad
   +
    \partial_x^2 B(x) H_{n}(x) a_n(t)
\nonumber
\\
& \qquad\qquad
   +
  \partial_x^3
      C H_{n}(x) a_n(t)
\\
&=
    \partial_x
    A(x) H_{n}(x) a_n(t)
\nonumber
\\
& \qquad\quad
   +
    \partial_x^2 
    B(x) H_{n}(x) a_n(t)
\nonumber
\\
& \qquad\qquad
   +
  \partial_x^2
      C
      \left(
        2 x H_{n} - H_{n+1}
      \right) 
      a_n(t)
\\
&=
    \partial_x
    A(x) H_{n}(x) a_n(t)
\nonumber
\\
& \qquad\quad
   +
    \partial_x^2 
  \left[
    B(x) 
   +
    2 x 
  \right]
  H_{n}(x) a_n(t)
\nonumber
\\
& \qquad\qquad
   +
  \partial_x^2
      C
      H_{n+1} a_n(t)
.
\end{align}

Note the appearance of the $H_{n+1}$.
Since we aim to reinstate the summation over all $H_{n}$, 
 we might reassign this to $B_{n+1}$, 
 so
~ 
\begin{align}
  A''_n(x)
&=
      A(x) 
\\
  B''_n(x) a_n(t)
&=
    B(x) a_n(t)
   +
    2 x C a_n(t)
   -
    \left( 1 - \delta_{0n} \right) a_{n-1}(t)
.
\end{align}

So we see no drift modification, 
 but a diffusion reduction instead.
However, 
 the $H_{n}$ equations are still cross-coupled, 
 since $B''_n a_n(t)$ has gained a dependence on $a_{n-1}$.
In the case of an even $n$
 the diffusion adjustment will depend on $n-1$ which is a 
 measure of the odd-ness of $P$ (as projected onto $H_{n-1}$); 
 the converse is true for odd $n$.
This has turned out to be just
 a re-representation of the effective-drift adjustment
 seen for $A'$ in the first method.


%
\subsection{Rebuilding the FPE}\label{S-thirdorder-rebuild}

Another (unsucessful) attempt to make a useful application
 of the special case
 in Sec. \ref{S-Hn}
 is to try to reinstate the summation over $n$,
 to convert eqn. (\ref{eqn-thirdderivs-hermiteFPE-n-trans})
 back into a true FPE.
Thus
~
\begin{align}
  \partial_t 
    \sum_n
      a_n(t) H_{n}(x)
&=
  \partial_x
  \left[
    A'_n(x)
   +
    \partial_x B'(x)
  \right]
  \sum_n
    a_n(t)
    H_{n}(x)
\\
  ? \Rightarrow
  \partial_t 
    P(x;t)
&=
  \partial_x
  \left[
    A'(x)
   +
    \partial_x B'(x)
  \right]
    P(x;t)
.
\label{eqn-thirdderivs-rebuilt-FPE}
\end{align}
Unfortunately this fails
 because $A'_n$ is dependent on $n$, 
 and we cannot reach the desire $n$-independent form for $A$
 seen in the target eqn. \eqref{eqn-thirdderivs-rebuilt-FPE}.
And even if I instead attempt to remove that $n$ dependence, 
 using eqn. (\ref{eqn-hermite-recurr1}), 
 i.e. $2n H_{n} =  2x H_{n+1} - H_{n+2}$, 
 I end up cross-coupling the $H_n$ contributions instead, 
 which is no better.


%


%
\section{Conclusion}

I have shown that in one specific case, 
 the Hermite polynomial recurrence relations
 can be used to develop an exact relationship
 between a constant third order derivative term in an FPE, 
 and the commonly understood diffusion and drift terms.
Although there seems no clear path to use this as a basis
 to solve FPE's with third order terms in general, 
 it can still be useful in applying check to an ongoing numerical solution
 to a truncated (standard) FPE equation.

%
\appendix

\section*{Note}

This is a previously unpublished fragment of my PhD research, 
 which I have dusted off and put here on the arXiv,
 in case someone finds it useful.

%
\bibliography{/home/physics/_work/bibtex.bib}

\end{document}